\documentclass[10pt,notitlepage,showpacs,pra,twocolumn]{revtex4-1}
\usepackage{ulem}
\usepackage{amsfonts}
\usepackage{amsmath}
\usepackage{amssymb}
\usepackage{graphicx}
\usepackage{float}
\usepackage[toc,page,header]{appendix}

\begin{document}

\title{Robust quantum switch with Rydberg excitations}
\author{Jing Qian$^{\dagger}$ }
\affiliation{Department of Physics, School of Physics and Material Science, East China
Normal University, Shanghai 200062, People's Republic of China}

\begin{abstract}
We develop an approach to realize a quantum switch for Rydberg excitation in atoms with $Y$-{\it typed} level configuration. We find that the steady population on two different Rydberg states can be reversibly exchanged in a controllable way by properly tuning the Rydberg-Rydberg interaction. Moreover, our numerical simulations verify that the switching scheme is robust against spontaneous decay, environmental disturbance, as well as the duration of operation on the interaction, and also a high switching efficiency is quite attainable, which makes it have potential applications in quantum information processing and other Rydberg-based quantum technologies.

%We develop a novel approach of switching excitations between two Rydberg states of an atom with a $Y$-{\it typed} level configuration. By turning on the intrastate Rydberg interaction of the strongly-coupled Rydberg state, the atom is counter-intuitively excited to the weakly-coupled Rydberg state, rather than the strongly-coupled state as in the case without the interaction. By investigating the relevant parameters we find the switching scheme is quite effective and robust, and it is insensitive to the strength of the intrastate interaction of the weakly-coupled state, the decay rate of the intermediate excited state, and the duration time of the switching process. Moreover, a switching cycle under realistic experimental conditions is theoretically investigated and a very high switching efficiency is confirmed by our numerical simulations. \textbf{Our switch scenario can serve as an efficient route to the selective excitations in multiple Rydberg states with a low demand of laser power, having practical applications in Rydberg entanglement preparation and quantum information processing}.
\end{abstract}
\email{jqian1982@gmail.com}
\maketitle
\preprint{}
%\begin{multicols}{2}

\section{Introduction}

Switch is a device that is capable of switching some kind of signals (e.g. current, voltage, energy, heat {\it et.al.}) between different pathways. Classical switch plays a vital role in electronics and signal processing. Extending such a concept into the quantum regime where the role of pathways is played by quantum states leads to the production of various quantum switches, 
such as the switchable acoustic meta-materials \cite{Wang14}, the current switch in quantum dots \cite{Brandes00,Dessotti16}, the superconducting switch \cite{Pechal16}, the fiber-optical switch \cite{Shea13} and so on.
In particular, for achieving an all-optical quantum switch, one promising way is coupling the atoms to a microscopic high-finesse cavity \cite{Davidovich93,Feng06,Chen13,Tiecke14} which can strongly enhance the light-atom interactions \cite{Haroche06}. Such a quantum optical switch has many promising applications, ranging from quantum information processing to quantum metrology \cite{Briegel98,Escartin06,Schirmer09,Qin16}.

Recently, Rydberg atoms have been manifested as an ideal candidate to study single-photon all-optical switches \cite{Escartin12,Baur14,Li15} and transistors \cite{Gorniaczyk14,Tiarks14,Gorniaczyk16}, mainly due to the presence of interatomic interactions \cite{Gallagher08,Marcassa14}. The ultra-strong interaction between two Rydberg states gives rise to blockade effect,  bringing on a strong enhancement for the light-atom interactions \cite{Heidemann07,Gaetan09,Pritchard10}. Moreover, the blockade effect can provide an efficient mechanism for controlling the quantum states of the atomic system itself. A simplest scheme can be carried out,
for example in a two-atom system, it prohibits the excitation of the second atom when the first one has already been excited to the Rydberg state. That is, it allows to control one atom's excitation or not via the status of the other \cite{Jaksch00,Lukin01,Comparat10}, achieving a switchable excitation between two atomic states.

In the present work we propose a new scheme of quantum switch based on two Rydberg atoms of same $Y$-\textit{typed} four-level configurations \cite{Liu15}. The special level configuration has two different Rydberg states: one is weakly coupled to the intermediated state and the other is strongly coupled. This enables two different excitation pathways labeled as ``OFF'' and ``ON'' by us [see Fig. \ref{modelandexc}(c)], which can be efficiently switched via the control of intrastate interaction of the strongly-coupled Rydberg state. 
The interstate interaction between different Rydberg states, as main disturbance for the status switch, is greatly suppressed by employing the feature of $nS$ Rydberg states that the strength of van der Waals (vdWs) interaction is not affected by Zeeman effect \cite{Walker08}.
The robustness of the switching scheme is confirmed by its low sensitivity to the other parameters of the system, such as the intra-state interaction of the weakly-coupled state, the decay rate of the intermediate state, and the duration time of the switching process. We present a detailed discussion of a realistic experimental implementation of the switch with $^{87}$Rb atoms and predict that the final switching efficiency will reach as high as 0.92.

%In ``switch-off" where the interaction $\mathcal{V}_{0,pp}$ to the strong coupling state is tuned to vanish, $\frac{1}{2}$ population is prepared in the single Rydberg state $\left\vert gp\right\rangle_{+}$. 
%While in ``switch-on" with $\mathcal{V}_{0,pp}\neq 0$ blocking the transition between single and double Rydberg states, almost $\frac{1}{2}$ population will transferred to the other single Rydberg state $\left\vert gs\right\rangle_{+}$. In other words, by operating the intrastate interactions $\mathcal{V}_{0,pp}$ to the strong coupling state, Rydberg excitation probability can be switched with a high efficiency between two pathways. 

\section{Model Description}.

\begin{figure}
\includegraphics[width=3.34in,height=3.3in]{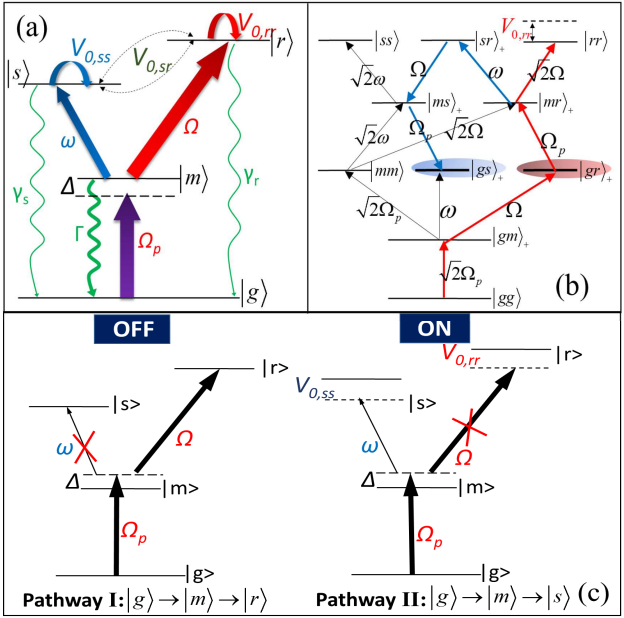}
\caption{(color online) (a) Level structure of a single atom we adopted with the detailed descriptions about level couplings in the main text. (b) Level structure of two interacting atoms with the interstate transitions and the corresponding Rabi frequencies marked out. 
(c) Two possible pathways for exciting atoms into Rydberg states $\left\vert s\right\rangle$ and $\left\vert r\right\rangle$. For $\Omega>\omega$ and the interaction $\mathcal{V}_{0,rr}=0$, the atom will be excited to the strongly-coupled state $\left\vert r\right\rangle$ through pathway $\rm{I}$, defined as the ``OFF'' status of the switch; if $\mathcal{V}_{0,rr}\neq 0$ the atom prefers the excitation to the weakly-coupled state $\left\vert s\right\rangle$ through pathway $\rm{II}$, defined as the ``ON'' status. For both cases the interaction strength $\mathcal{V}_{0,sr}=0$ and $\mathcal{V}_{0,ss}$ is arbitrary.}
\label{modelandexc}
\end{figure}
 
Our model consists of two identical Rydberg atoms in frozen-gas limit. As presented in Fig. \ref{modelandexc}(a), each atom has a $Y$-\textit{typed} level structure that the ground state $%
\left\vert g\right\rangle $ is coupled to the middle state $\left\vert
m\right\rangle $ via a laser field with Rabi frequency $\Omega_{p} $ and detuning $\Delta$, and $%
\left\vert m\right\rangle $ is further resonantly coupled to two different Rydberg states $%
\left\vert s\right\rangle $ and $\left\vert r\right\rangle $ with Rabi
frequencies $\omega $ and $\Omega $, respectively. The Hamiltonian for a single atom $k$ reads ($\hbar=1$ everywhere)
\begin{equation}
\mathcal{H}_{k}=\Delta \sigma_{mm}^{(k)}+(\Omega_{p}\sigma_{gm}^{(k)}+\Omega\sigma_{mr}^{(k)}+\omega\sigma_{ms}^{(k)}+\text{H.c.}),
 \label{Sham}
\end{equation}
where the atomic operators $%
\sigma _{\alpha \beta }^{\left( k\right) }=\left\vert \alpha
_{k}\right\rangle \left\langle \beta _{k}\right\vert $, $\alpha,\beta\in \{g,m,s,r\}$.

The properties of the interaction between atoms are dependent on the Rydberg states we chose. For $nS$ Rydberg states, in the absence of electrostatic field the interaction is dominant by the second-order dipole-dipole interaction (i.e. vdWs interaction) \cite{Saffman10}. It can be further classified as (i) the intrastate interaction $\mathcal{V}_{ss\left( rr\right) }=\mathcal{V}_{0,ss\left( rr\right)
}\left\vert ss\left( rr\right) \right\rangle \left\langle ss\left( rr\right)
\right\vert $ with the strength $\mathcal{V}_{0,ss\left( rr\right) }=C_{6}^{s\left( r\right) }/R^{6}$, which present if both atoms settle in a same Rydberg state $\left\vert s\right\rangle $ or $\left\vert r\right\rangle $, (ii) the interstate interaction $\mathcal{V}_{sr}=\mathcal{V}_{0,sr}\left( \left\vert
sr\right\rangle \left\langle rs\right\vert +\left\vert rs\right\rangle
\left\langle sr\right\vert \right) $ for atoms
in different Rydberg states with $\mathcal{V}_{0,sr}=C_{6}^{sr}/R^{6}$. Here $R$ represents the separation between atoms, $C_6^{s,r,sr}$ are the interaction coefficients, and $\left\vert
\alpha \beta \right\rangle \equiv \left\vert \alpha \right\rangle_1 \otimes
\left\vert \beta \right\rangle_2 $ are the two-atom states. Then the total Hamiltonian $\mathcal{H}$ is obtained from the sum of $\mathcal{H}_{k=1,2}$ and the interaction terms,
\begin{equation}
\mathcal{H=H}_{1}+\mathcal{H}_{2}+\mathcal{V}_{ss}+\mathcal{V}_{rr}+\mathcal{V}_{sr}.
\label{Hamtot}
\end{equation}

For a single atom, there exist two different pathways for excitation, \rm{I}: $\left\vert g\right\rangle \rightarrow \left\vert m\right\rangle \rightarrow \left\vert
r\right\rangle $ and \rm{II}: $\left\vert
g\right\rangle \rightarrow \left\vert m\right\rangle \rightarrow \left\vert
s\right\rangle $. We consider the condition $\Omega >\omega $ so state $\left\vert r\right\rangle $ is the strongly-coupled state and the excitation pathway $\rm{I}$ is preferred, while $\left\vert s\right\rangle $ is the weakly-coupled state and the pathway $\rm{II}$ is less taken. 
Considering the long lifetime of states $\left\vert s\right\rangle $ and $\left\vert r\right\rangle $, their spontaneous decays $\gamma_{s} $ and $\gamma_{r}$ are far less than the decay $\Gamma$ for state $\left\vert m\right\rangle $.
For simplicity, we first assume $\Omega_p=\Omega$, $\Delta=0$ and $\gamma_s=\gamma_r=\gamma$. As shown in Fig. \ref{modelandexc}(c), if $\Omega \gg \omega $, it is easy to envision that there is a steady state that almost $1/2$ population transfers from $\left\vert g\right\rangle $ to $\left\vert r\right\rangle $ through pathway \rm{I} with no population on $\left\vert m\right\rangle $ or  $\left\vert s\right\rangle $, which is labeled as ``OFF'' state. However, we will show that considerable population would counter-intuitively transfer into the weakly-coupled state $\left\vert s\right\rangle $ through pathway $\rm{II}$ once the interaction $\mathcal{V}_{0,rr}\neq 0$. This process is found to be fully irrespective of the exact interaction strength $\mathcal{V}_{0,ss}$ and can serve as a controllable switch between the two status ``OFF'' and ``ON'' corresponding to different Rydberg excitations. 

\section{Single-atom case}
\begin{figure}
\includegraphics[width=2.54in,height=1.8in]{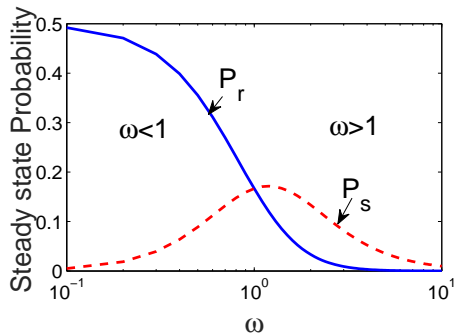}
\caption{(Color online) For the single-atom case, the steady probabilities of Rydberg state $\left\vert r\right\rangle $ (blue solid) and $\left\vert s\right\rangle $ (red dashed) are plotted as functions of $\omega$ with the decays $\gamma=0.001$ and $\Gamma=1.0$. All frequencies are scaled by $\Omega$.}
\label{Extnoint}
\end{figure}

We begin with the status ``OFF'' which can be analyzed in single-atom frame due to the absence of Rydberg interaction. The analytical expression for steady state can be obtained by solving the master equation $%
\Dot{\rho}_{k}=-i\left[ \mathcal{H}_{k},\rho_{k} \right] +\mathcal{L}_{k}\left[ \rho_{k} %
\right] $ ($k=1,2$) with $\rho_{k}$ and $\mathcal{H}_{k}$ the single-atom density matrix and Hamiltonian, respectively. Here the Lindblad superoperator $\mathcal{L}_{k}[\rho]$ is given by
\begin{widetext}
\begin{equation}
%\begin{split}
\mathcal{L}_{k}\left[ \rho \right] =\Gamma
\left( \sigma _{gm}^{(k)}\rho_{k} \sigma _{mg}^{(k)}-\frac{\left\{ \sigma _{mm}^{(k)},\rho_{k}
\right\}}{2}\right) +\gamma \left( \sigma _{gs}^{(k)}\rho_{k} \sigma _{sg}^{(k)}-\frac{
\left\{ \sigma _{ss}^{(k)},\rho_{k} \right\}}{2} \right) 
+\gamma \left( \sigma
_{gr}^{(k)}\rho_{k} \sigma _{rg}^{(k)}-\frac{\left\{ \sigma _{rr}^{(k)},\rho_{k} \right\}}{2} \right),
%\end{split}
\end{equation}
\end{widetext}
which describes the effect of spontaneous decays from states $\left\vert m\right\rangle$, $\left\vert s\right\rangle$, and $\left\vert r\right\rangle$. 
In the following calculations, we use $\Omega$ ($\Omega^{-1}$) as the frequency (time) unit, leading to normalized parameters as $\Omega_{p}\to\Omega_{p}/\Omega$, $\omega\to\omega/\Omega$, $\Gamma\to\Gamma/\Omega$, $\gamma\to\gamma/\Omega$, $\Delta\to\Delta/\Omega$, $\mathcal{V}_{0,rr(ss)}\to\mathcal{V}_{0,rr(ss)}/\Omega$, $\mathcal{V}_{0,sr}\to\mathcal{V}_{0,sr}/\Omega$, and  $t\to \Omega t$.
Then the steady population of $\left\vert r\right\rangle$ is
%\begin{widetext}
\begin{equation}
%\begin{aligned}
 P_{r} = \frac{ 1}{\frac{ 4\Gamma
(1+\omega ^{2})+\gamma(8+\Gamma^{2})
}{16}\frac{4+\gamma(\Gamma+\gamma)}{\Gamma+\gamma}
+\frac{(1+\omega ^{2}) \Gamma\gamma+4(1+\omega ^{2})^{2} }{4} }  \label{pstatep}
%\end{aligned}
\end{equation}
%\end{widetext}%
and of $\left\vert s\right\rangle $ is $P_{s}=\omega ^{2}P_{r}
$. %satisfying $P_{s}<P_{r}$ due to $\omega<1 $. 
In the limit of $\omega\ll1$, Eq. (\ref{pstatep}) reduces to
$P_{r} \to 1/2$ and $P_{s}\rightarrow 0$, coinciding with our previous predictions about the status ``OFF''.
In Fig. \ref{Extnoint}, we plot $P_{r}$ and $P_{s}$ as functions of $\omega$, which shows that $P_{r}$ decays and $P_{s}$ grows up as $\omega$ increases and they become equal at $\omega =1$. For further increased $\omega$, both of them decrease but at different rates. 
The monotonous decrease of $P_{r}$ is easy to understand, while the variation of $P_{s}$ is ascribed to the electromagnetically induced transparency (EIT) effect in pathway II. A unique feature of the effect is that the excitation probability decreases as enhancing the coupling laser strength \cite{Garttner14}.

%For comparison, we also numerically study the other case with two noninteracting atoms and plot the single atom excitation probability $P_{gp_+}$ (blue dashed curve) of state $\left\vert gp\right\rangle_{+} $, $P_{gs_+}$ (red dashed curve) of state $\left\vert gs\right\rangle_{+} $ in Fig. \ref{Extnoint}. Here, the single atom excitation probability means the probability of exciting one of the two atoms to the Rydberg state $\left\vert p\right\rangle $
%or $\left\vert s\right\rangle $, and the other atom still stays in $\left\vert g\right\rangle $. As Fig. \ref{Extnoint} displays, $P_{gp_+}$ ($P_{gs_+}$) is slightly larger than $P_{p}$ ($P_{s}$), but they two share same tendency. Especially, in the limits of $\omega \ll \Omega $ and $\Omega \ll \omega $ where the $Y$-\textit{type} scheme reduces into a three-level Ladder scheme, they are exactly same since only one of the two transitions play roles then.

\section{Two-atom Case}

Turning to the picture of two interacting atoms, if the initial state is $|gg\rangle$ the total Hamiltonian $\mathcal{H}$ can be expanded by the ten symmetric two-atom bases only, $\{\left\vert gg\right\rangle$, $\left\vert mm\right\rangle$, $\left\vert ss\right\rangle$, $\left\vert rr\right\rangle$, $\left\vert gm\right\rangle_{+}$, $\left\vert gs\right\rangle_{+}$, $\left\vert gr\right\rangle_{+}$, $\left\vert ms\right\rangle_{+}$, $\left\vert mr\right\rangle_{+}$, $\left\vert sr\right\rangle_{+}\}$ where $\left\vert \alpha\beta\right\rangle_{\pm}=(\left\vert  \alpha\beta\right\rangle\pm\left\vert  \beta\alpha\right\rangle)/\sqrt{2}$, with the asymmetric states $\left\vert \alpha\beta\right\rangle_{-}$ safely ignored \cite{Moller08}.
The coupling strategie and strength among them are presented in Fig. \ref{modelandexc}(b). 
In the absence of all Rydberg interaction, the population transfer is mainly following the approach $\left\vert  gg\right\rangle$$\to$$\left\vert  gm\right\rangle_+$$\to$$\left\vert  gr\right\rangle_+$$\to$$\left\vert  mr\right\rangle_+$$\to$$\left\vert  rr\right\rangle$ (red arrows) due to the stronger coupling strengths ($\propto\Omega$). Finally, the long-lived states $\left\vert  gg\right\rangle$, $\left\vert  gr\right\rangle_+$ and $\left\vert  rr\right\rangle$ are stably populated. Less population is found to accumulate in middle states $\left\vert  gm\right\rangle_+$ and $\left\vert  mr\right\rangle_+$ for their short lifetimes, and in $\left\vert  mm\right\rangle$ and $\left\vert  gs\right\rangle_+$ for large decay rate $\Gamma$ and small coupling strength $\omega$, respectively. The  steady populations are $P_{gr_+}\approx0.5$ and $P_{gs_+}\approx 0$ which is same as the status ``OFF'' analyzed in the single-atom case. Note that $P_{gg}+P_{rr}=1-P_{gr_+}\approx0.5$.
Once the intrastate interaction $\mathcal{V}_{0,rr}$ is nonzero, giving rise to an energy shift on state $\left\vert  rr\right\rangle$, the transition from $\left\vert  mr\right\rangle_+$ to $\left\vert  rr\right\rangle$ will be affected. If the condition for strong blockade, $\mathcal{V}_{0,rr}>\sqrt{2}$, is satisfied \cite{Gaetan09}, the transition to the doubly Rydberg excited state will be fully suppressed. Instead, the population moves towards $\left\vert  sr\right\rangle_+$ when the interstate interaction $\mathcal{V}_{0,sr}=0$, 
which leads to the second transition pathway $\left\vert  mr\right\rangle_+$$\to$$\left\vert  sr\right\rangle_+$$\to$$\left\vert  ms\right\rangle_+$$\to$$\left\vert  gs\right\rangle_+$ (blue arrows). Finally the steady populations are  $P_{gr_+}\approx0$ and $P_{gs_+}\approx 0.5$, corresponding to the status ``ON'' for our quantum switch.  $P_{sr_{+}}$ is also dominantly occupied besides $P_{gs_+}$ and $P_{gg}\approx0$.

%Hence, it is possible to realize a switchable Rydberg excitation via the control of $\mathcal{V}_{0,rr}$ only.

\begin{figure}
\includegraphics[width=3.45in,height=1.55in]{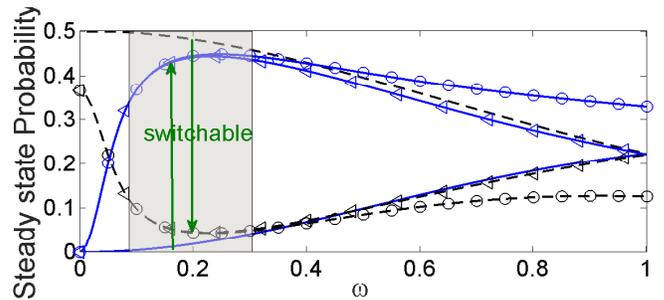}
\caption{(color online) Steady probabilities $P_{gs_+}$ (blue) and $P_{gr_+}$ (black) as functions of $\omega$ with different intrastate interactions $\mathcal{V}_{0,rr}$ and $\mathcal{V}_{0,ss}$. The interstate interaction $\mathcal{V}_{0,sr}$ is kept at zero. The shaded area of $\omega\in(0.1,0.3)$ is suitable for switching. Relevant parameters are described in the main text.}
\label{excitationP}
\end{figure}

The above qualitative analysis have been verified by numerically solving the master equation of two atoms, $\Dot{\rho}=-i\left[ \mathcal{H},\rho \right] +\mathcal{L}_{1}\left[ \rho \right] +\mathcal{L}_{2}\left[ \rho \right]$, in which $\rho$ is replaced by a two-atom density matrix.
The steady populations $P_{gr_+}$ and $P_{gs_+}$ for states $\left\vert gr\right\rangle_{+}$ and $\left\vert gs\right\rangle_{+}$ are illustrated as functions of $\omega$ in Fig. \ref{excitationP} for  three different cases: (i) $\mathcal{V}_{0,rr}=\mathcal{V}_{0,ss}=0$ (the black dashed curve for $P_{gr_+}$ and the blue solid curve for $P_{gs_+}$), (ii) $\mathcal{V}_{0,rr}=1.0$ and $\mathcal{V}_{0,ss}=0$ (the black dashed curve with circles for $P_{gr_+}$ and the blue solid curve with circles for $P_{gs_+}$), and (iii) $\mathcal{V}_{0,rr}=1.0$ and $\mathcal{V}_{0,ss}=1.0$ (the black dashed curve with triangles for $P_{gr_+}$ and the blue solid curve with triangles for $P_{gs_+}$). In case (i) we find $P_{gr_+}\approx0.5$ and $P_{gs_+}\approx0.0$ at $\omega\ll 1$ and they two become equal as $\omega$ increases to $1$, same as in Fig. \ref{Extnoint} obtained in the single-atom frame. When the intrastate interaction $\mathcal{V}_{0,rr}$ is present, see the cases (ii) and (iii), there is a counterintuitive reversal of $P_{gr_+}$ and $P_{gs_+}$ at $\omega\gtrsim0.05$, indicating a large fraction of population transferred from $\left\vert  gr\right\rangle_+$ to $\left\vert  gs\right\rangle_+$. In the shadow region of $0.1\lesssim \omega\lesssim 0.3$, $P_{gs_+}$ and $P_{gr_+}$ attain peak and off-peak values, respectively. Especially, by comparing cases (ii) and (iii) we find their variations are quite insensitive to the interaction strength $\mathcal{V}_{0,ss}$ in this region, which is ideally suited for operating the quantum switch.

\section{The Switch Efficiency}
To investigate the performance of the quantum switch, we first define the switching efficiency as
\begin{equation}
 \eta=\frac{P_{gs_+}^{on}}{P_{gr_+}^{off}},
\end{equation}
where $P_{gr_+}^{off}$ and $P_{gs_+}^{on}$ are the steady population of $\left\vert gr\right\rangle_+$ in status ``OFF'' and of $\left\vert gs\right\rangle_+$ in status ``ON', respectively. The status is switched by turning up or down the interaction $\mathcal{V}_{0,rr}$. For a ideal switch the population $P_{gs_+}^{on}=P_{gr_+}^{off}=0.5$ and the efficiency $\eta=1$.

\begin{figure}
\includegraphics[width=3.4in,height=1.1in]{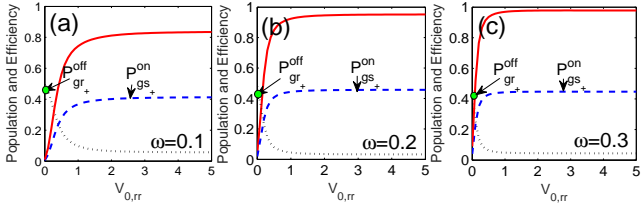}
\caption{(Color online) Steady populations $P_{gr_+}$ (black dotted) and $P_{gs_+}$ (blue dashed), and the switch efficiency $\eta$ (red solid) versus the interaction strength $\mathcal{V}_{0,rr}$ in the cases of (a) $\omega=0.1$, (b) $\omega=0.2$ and (c) $\omega=0.3$. $\omega$ is normalized by the Rabi frequency $\Omega$. $P_{gr_+}^{off}$ is the value of $P_{gr_+}$ at $\mathcal{V}_{0,rr}=0$, marked by the green point, and $P_{gs_+}^{on}$ is same as $P_{gs_+}$.}
\label{efficiency}
\end{figure}

In Fig. \ref{efficiency}(a-c) we show the dependence of $\eta$ on the interaction strength $\mathcal{V}_{0,rr}$ under the different relative couplings $\omega$ ($\omega$ is already normalized by $\Omega$). For comparison, the steady populations $P_{gs_+}$ (blue dashed) and $P_{gr_+}$ (black dotted) are presented in the same frame. $\eta$ reaches a saturation value and no longer changes with $\mathcal{V}_{0,rr}$ once $\mathcal{V}_{0,rr}>\sqrt{2}$, satisfying the two-atom strong blockade condition \cite{Gaetan09}. This brings us a big advantage at selections of state $\left\vert r\right\rangle$ in practice, especially for the atoms with multiple Rydberg energy levels. 
Besides, the saturation value of $\eta$ is observed to be enhanced with the increase of $\omega$, which is attributed to the slight changes of $P_{gr_+}^{off}$ (green dot) and $P_{gs_+}^{on}$ (blue dashed curve).  For instance, in the case of $\omega=0.3$ the saturated $\eta\to0.978$.
In the opposite case of $\mathcal{V}_{0,rr}<\sqrt{2}$, $\eta$ and $P_{gs_+}^{on}$ rapidly falls to zero with the decrease of $\mathcal{V}_{0,rr}$.

\begin{figure}
\includegraphics[width=3.4in,height=2.5in]{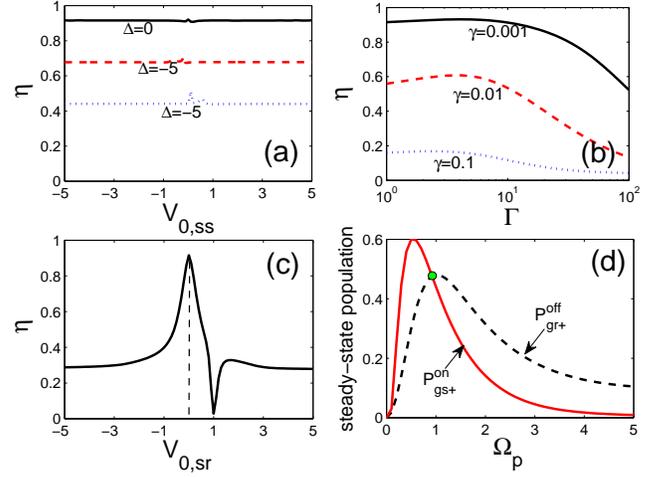}
\caption{(Color online) The switching efficiency $\eta$ as a function of (a) the intrastate interaction $\mathcal{V}_{0,ss}$ for $\Delta=0,-5,5$ and the other parameters $\Gamma=1.0$, $\gamma=0.001$, $\mathcal{V}_{0,sr}=0$; (b) the decay $\Gamma$ from intermediate state for the Rydberg-state decay $\gamma=0.001,0.01,0.1$ and the other parameters $\mathcal{V}_{0,ss}=1.0$, $\mathcal{V}_{0,sr}=0$, $\Delta=0$; (c) the interstate exchange interaction $\mathcal{V}_{0,sr}$ for the other parameters $\mathcal{V}_{0,ss}=1.0$, $\Gamma=1.0$, $\gamma=0.001$, $\Delta=0$. Additionally, $\omega=0.2$ and $\mathcal{V}_{0,rr}=1.0$. The steady-state population $P_{gr+}^{off}$ and $P_{gs+}^{on}$ as a function of the driving laser $\Omega_p$ are shown in (d). All frequencies are scaled by $\Omega$.}
\label{efficiency2}
\end{figure}

Except for $\mathcal{V}_{0,rr}$ and $\omega$, we also explore the influence of  other parameters on the switching scheme, including the interaction $\mathcal{V}_{0,ss}$ of the weakly-coupled Rydberg state, the spontaneous decay $\Gamma$ of the middle state, and the Rabi frequency $\Omega_p$. 
In Fig. \ref{efficiency2}(a) we find that $\eta$ keeps constant with $\mathcal{V}_{0,ss}$. This is due to the isolation of state $\left\vert  ss\right\rangle$ from the transfer pathway as presented in Fig. \ref{modelandexc}(b), so that the energy shift of $\left\vert  ss\right\rangle$ induced by interaction $\mathcal{V}_{0,ss}$ has no effect on $\eta$. However, an off-resonant transition $\left\vert  g\right\rangle\to\left\vert  m\right\rangle$ characterized by detuning $\Delta$ will reduce the steady population of Rydberg state, resulting in a decrease of $\eta$. We find that the switching scheme is robust to the middle-state decay $\Gamma$.  As displayed in Fig. \ref{efficiency2}(b), $\eta$ keeps almost unvaried in a  broad regime of $1.0<\Gamma<10$ and start to slowly decrease only when $\Gamma>10$. In contrast, a larger decay $\gamma$ means a quick decay from Rydberg states, which directly causes a drop of $\eta$.

In addition to the parameters above, it should be stressed that the interstate interaction $\mathcal{V}_{0,sr}$ strongly destroys the switching efficiency. As shown in Fig. \ref{efficiency2}(c), $\eta$ rapidly decreases as long as $|\mathcal{V}_{0,sr}|\neq0$. That is because a nonzero $\mathcal{V}_{0,sr}$ will shift level $\left\vert sr\right\rangle_+$ and hinder the transition from $\left\vert  mr\right\rangle_+$ to $\left\vert  sr\right\rangle_+$, see Fig. \ref{modelandexc}(b). Worse, the competition between transitions of $\left\vert  mr\right\rangle_+\to\left\vert  rr\right\rangle_+$ and $\left\vert  mr\right\rangle_+\to\left\vert  sr\right\rangle_+$ is most serious when $\mathcal{V}_{0,sr}=\mathcal{V}_{0,rr}$, resulting in a near zero $\eta$.  
So the suppression of $\mathcal{V}_{0,sr}$ is a crucial condition for our approach of quantum switching. It can be guaranteed if we choose two $nS$ states with large difference in principle quantum numbers $n$ as Rydberg states. In the absence of applied electrostatic fields, the interstate interactions are negligible compared with the intrastate interactions, which has been confirmed theoretically \cite{Olmos11,Cano14} as well as experimentally \cite{Teixeira15}. 

Finally, we consider a more general case $\Omega_p\neq1.0$ (i.e. $\Omega_p\neq\Omega$). Since the common limit $P_{gs_+}^{on}=P_{gr_+}^{off}=0.5$ is unable to maintain in this case, the definition of $\eta$ is no longer rigorous. We then show $P_{gr+}^{off}$ and $P_{gs+}^{on}$ versus $\Omega_p$ individually in Fig. \ref{efficiency2}(d). As the increase of $\Omega_p$, $P_{gr+}^{off}$ (black dashed) exhibits a clear reduction after reaching its maximum value 0.5 at $\Omega_p=1.0$. Similar trends are observed in $P_{gs+}^{on}$ (red solid) but the maximum 0.6 appears at $\Omega_{p}\approx0.6$. The reductions are because that for a large $\Omega_p$ the transition of $\left\vert  gm\right\rangle_+\to\left\vert  mm\right\rangle$ is enhanced which results in a decrease of excitations to $\left\vert  gs\right\rangle_+$ and $\left\vert  gr\right\rangle_+$, see Fig. \ref{modelandexc}(b). Hence, we conclude that $\Omega_p=1.0$ is an optimized value for our switching scheme, because the population of status ``OFF'' and ``ON'' are asymmetry for other values.

\section{Experimental Implementation}
After carefully researching the steady state of the switching system, we now turn to study the switching dynamics by numerically simulation with a series of practical experimental parameters. We assume two $^{87}$Rb atoms are respectively confined in two independent optical dipole traps whose separation $R$ can be adjusted from $15\mu$m to $4.0\mu$m by changing the incidence angle of the optical beams in a duration $\tau$ of the orders of several $\mu s$ \cite{Gaetan09,Beguin13}. For the atomic states $\left\vert g\right\rangle=\left\vert 5s_{1/2}\right\rangle$, $\left\vert m\right\rangle=\left\vert 5p_{3/2}\right\rangle$, and Rydberg $nS$ states $\left\vert s\right\rangle=\left\vert 47s\right\rangle$, $\left\vert r\right\rangle=\left\vert 65s\right\rangle$, the vdWs intrastate interaction coefficients are $C_6^{r}/2\pi=50.4$GHz$\mu$m$^6$ and $C_6^{s}/2\pi=1.0$GHz$\mu$m$^6$, and the spontaneous decay rates are $\Gamma/2\pi=6.1$MHz, $\gamma_s/2\pi=7$kHz, and $\gamma_r/2\pi=3$kHz (the effective lifetime is approximately 140$\mu$s and 320$\mu$s for $\left\vert 47s\right\rangle$ and $\left\vert 65s\right\rangle$, respectively, at 50 $\mu$K) \cite{Beterov09}. 
In our switching operation, the initial separation is $R$=15$\mu$m, leading to the interaction strength $\mathcal{V}_{0,rr}^{off}/2\pi=0.004$MHz and $\mathcal{V}_{0,ss}^{off}/2\pi=8\times 10^{-5}$MHz. When $R$ is reduced to 4.0$\mu$m, the interactions are enhanced to $\mathcal{V}_{0,rr}^{on}/2\pi=12.3$MHz and $\mathcal{V}_{0,ss}^{on}/2\pi=0.24$MHz. For Rydberg states $\left\vert 47s\right\rangle$ and $\left\vert 65s\right\rangle$, we have $C_{6}^{r}\gg C_{6}^{s}\gg C_{6}^{sr}$ due to the large difference in principle quantum numbers of the two Rydberg states \cite{Levi15}, so that the interstate interaction $\mathcal{V}_{0,sr}$ is largely suppressed and can be safely neglected. The Rabi frequencies, $\Omega/2\pi=10$MHz and $\omega/2\pi=2$MHz, are typical of current experiments.

\begin{figure}
\includegraphics[width=3.45in,height=3.25in]{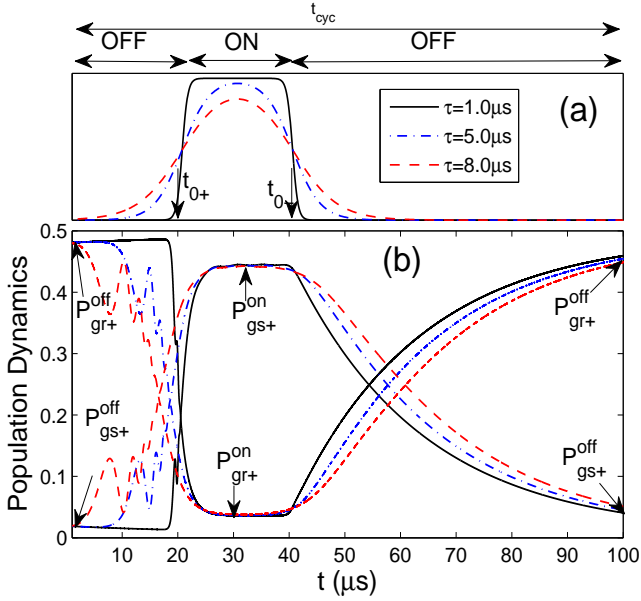}
\caption{(Color online) (a) Time dependence of interaction $V_{0,rr}(t)$ in the switching process for different optical switching durations $\tau=1.0\mu$s (black solid), $5.0\mu$s (blue dash-dotted) and $8.0\mu$s (red dashed); (b) The dynamical evolution of the population $P_{gr_+}(t)$ and $P_{gs_+}(t)$ under the sequence of $V_{0,rr}(t)$ in a normal and a reverse order. The parameters adopted in the simulation are described in the main text.
}
\label{response}
\end{figure}

To simulate the variation of the interaction under control, we introduce a time-dependent pulse sequence of $\mathcal{V}_{0,rr(ss)}(t)$ for a complete switching cycle consisting of three status: OFF, ON, and OFF,
\begin{widetext}
\begin{equation}
\mathcal{V}_{0,rr(ss)}(t) = \frac{\mathcal{V}_{0,rr(ss)}^{on}-\mathcal{V}_{0,rr(ss)}^{off}}{4}\left [1+\tanh(\frac{t-t_{0_+}}{\tau})\right]\times\left[1-\tanh(\frac{t-t_{0_-}}{\tau}) \right]
\end{equation}
\end{widetext}
where $\mathcal{V}_{0,rr(ss)}^{on}$ and $\mathcal{V}_{0,rr(ss)}^{off}$ take the previously estimated values of interaction strength for status ``ON'' and ``OFF'', respectively, $\tau$ is the switching duration characterizing the changing speed of the interaction, and $t_{0_+}$ and $t_{0_-}$ are the critical switching moments of $\mathcal{V}_{0,rr}^{off}\to\mathcal{V}_{0,rr}^{on}$ and $\mathcal{V}_{0,rr}^{on}\to\mathcal{V}_{0,rr}^{off}$, respectively. As shown in Fig. \ref{response}(a), the larger the $\tau$ is, the slower and smoother the switching operation between $\mathcal{V}_{0,rr}^{off}$ and $\mathcal{V}_{0,rr}^{on}$ is. The total duration of the switching cycle is $100\mu s$ which is less than the lifetime of the Rydberg states. 
As discussed before, the excitation is independent on the interaction $\mathcal{V}_{0,ss}(t)$, so for simplicity we assume it has a similar tendency of change with $\mathcal{V}_{0,rr}(t)$ here.

The dynamical evolution of the population $P_{gr_+}(t)$ and $P_{gs_+}(t)$, in response to the variation of interaction $\mathcal{V}_{0,rr(ss)}$, are obtained by numerically solving the master equation and displayed in Fig. \ref{response}(b).
Two conversions are clearly present at $t_{0_+}=20\mu$s and $t_{0_-}=40\mu$s. When $0<t<t_{0_+}$ the status ``OFF'' with the situation $P_{gr_+}^{off}\gg P_{gs_+}^{off}$ is maintained. A fast exchange of the population occurs around $t_{0_+}$ due to the switch $\mathcal{V}_{0,rr}^{off}\to\mathcal{V}_{0,rr}^{on}$, leading to the status ``ON'' with the reversed situation $P_{gs_+}^{on}\gg P_{gr_+}^{on}$ during the following period of time $t_{0_+}<t<t_{0_-}$. The second conversion takes place around $t_{0_-}$ when the interaction $\mathcal{V}_{0,rr}(t)$ is tuned down again, but at this time a longer period is required for the retrieval of the population for status ``OFF'' due to the different transition pathway.
The significant oscillation of population appears only at the period $0<t<t_{0_+}$ with a larger duration $\tau$. This is because in status ``OFF'' except $\left\vert gr_{+}\right\rangle$, $\left\vert gg\right\rangle$ and $\left\vert rr\right\rangle$ are also stably occupied, the resonant excitation between $\left\vert gg\right\rangle$ and $\left\vert gr_{+}\right\rangle$ will give rise to a Rabi-like oscillation if the duration of the switch is long enough. As an opposite example, in status ``ON'' only $\left\vert sr_{+}\right\rangle$ is dominantly occupied except for $\left\vert gs_{+}\right\rangle$, so the resonant excitation between them is not isolated but suffers from a strong decoherence. Hence, when $t>t_{0_-}$, even if $\tau$ is large there is no oscillation but a smooth redistribution of the population via decay process, which requests a longer time determined by the lifetime of Rydberg states.  
Based on our numerical simulations with practical parameters, a realistic switching efficiency is estimated as $\eta=P_{gs+}^{on}/P_{gr+}^{off}\approx0.92$ with the total operating time $t_{cyc}=100\mu$s.

\section{Applications and Conclusions}

The quantum switch we present here based on the controllable strong interaction between two Rydberg atoms, but different from the single-photon transistor with Rydberg blockade \cite{Gorniaczyk14}, it enables an efficient and compact transition between two symmetric singly Rydberg excited states $|gr\rangle_+$ and $|gs\rangle_+$. With appropriate applications and developments, this will broaden exciting perspectives on quantum information processing with Rydberg atoms. For example, owing to its long lifetime and entanglement \cite{Johnson10}, the singly excited states can become an excellent carrier of quantum information. Then the reversible and swift switch of these states is a requisite operation for implementation of information transfer and quantum computation. 
Especially, the considerable separation ($\sim 10\mu m$) between two Rydberg atoms in our design allows local operations on one of them individually, served with our switching on two-atom states, various quantum logic gates are hopefully realizable \cite{Xia13,Maller15,Beterov16,Theis16}. Besides, the Rydberg atomic pair-state interferometer has been experimentally realized \cite{Nipper12} recently. A high-precision quantum switch between different Rydberg excitations can enrich its measurement objects and develop the application of Rydberg atoms in quantum metrology. Finally, Rydberg dressing has been proposed to realize a number of interesting phases in ultra-cold gases, such as rotons and solitons \cite{Henkel10,Pupillo10}. An extension of our switch in a many-atom case will allow a more complex structure of Rydberg dressing, which makes it possible to simulate various and exotic spin-dependent phases by Rydberg atoms \cite{Weimer10}.

%The work presented here serves as a new type of particular switch with Rydberg atoms, bringing more perspectives on coherent quantum state engineering using Rydberg atoms. First, the excitation to a single specified level with large principle quantum number $n$ requires the lasers with short pulse or strong power \cite{Kozak13,Hart16}, which adds difficulty to the experimental practice. Based on the idea of our scheme, the required excitation laser to the weakly-coupled Rydberg state can save at least one order of magnitude by transferring population from another strongly-coupled channel. Second, by preparing atoms in two microtraps separated by a few $\mu$m (see Experimental Implementation), our scheme may allow further studies of transferring excitation between two atomic traps or large arrays of Rydberg atoms, implementing selective manipulation of individual atoms in a larger ensemble \cite{Labuhn14} or exploring the effect of Rydberg interactions at resonant pair states \cite{Reinhard08}. Third, as we know deterministic manipulation of entanglement with Rydberg states induced by strong blockade has been widely investigated in two \cite{Gaetan09} or more atoms \cite{Stanojevic09}. Here, if the implementation works in a time that is smaller than the Rydberg-state lifetime, the quantum switch mechanism can enable the creation of a coherent superposition of one ground state $\left\vert g\right\rangle$ and one Rydberg state $\left\vert s\right\rangle$ or $\left\vert r\right\rangle$, achieving switchable operation for Rydberg entanglements within a few $\mu$s. 

To conclude, our work presents a robust and experimentally feasible scheme of quantum switch, implemented in a system of two interacting Rydberg atoms. Each atom has a $Y$-\textit{typed} level structure with two highly-excited Rydberg states. We show that which Rydberg state to be excited can be simply and effectively controlled by opening or closing the intrastate interaction of the strongly-coupled Rydberg state. 
After systematically investigating the steady state and the dynamics of the system in a numerical way, we verify the robustness of the scheme by presenting its insensitivity to the self-interaction of the weakly-coupled Rydberg state, the decay of intermediate state, and the duration time for switching. Our method is suitable for two Rydberg $nS$ states in which the interstate exchange interaction between them can be totally suppressed by considering two $nS$ states with large different principle quantum numbers. More possibilities for the implementation with other energy levels may work, e.g. by applying an external electrostatic field \cite{Petrosyan14}. We show a numerical simulation of switch operation in $^{87}$Rb atoms under realistic experimental conditions and find the switch efficiency approaching as high as 0.92. A many-atom case maybe treated as a good extension to the current scheme in the future, requiring more attentions to complex energy levels and transitions. We also plan to develop the applications of such particular switch in the fields of quantum information processing and other quantum devices.

J. Qian thanks J. F. Chen for providing useful timescales (a few microseconds) of adjusting the inter-atomic separation $R$  in experiment. This work is supported by the NSFC under Grants No. 11474094, No. 11104076, the
Specialized Research Fund for the Doctoral Program of Higher Education No.
20110076120004.

%\end{multicols}{2}
\end{document}